\documentclass[prb,amsmath,amssymb,superscriptaddress,twocolumn]{revtex4}
\usepackage{float}
\usepackage{amsmath}
\usepackage{amssymb}
\usepackage{amsfonts}
\usepackage{euscript}
\usepackage{enumerate}
\usepackage{hhline}
\usepackage{pslatex}
\usepackage{tabularx}
\usepackage[usenames,dvipsnames]{xcolor}

\usepackage{graphicx}

\usepackage{dcolumn}
\usepackage{bm}
\usepackage[sort&compress]{natbib}

\usepackage{lipsum}

\makeatletter
\renewcommand{\@biblabel}[1]{#1. }
\renewcommand{\@dotsep}{500}
\renewcommand{\@pnumwidth}{0em}
\renewcommand{\l@figure}[2]{
\@dottedtocline{1}{1.5em}{2em}{Figure #1}{}\vspace{15pt}}

\newcommand{\ks}[1]{\textcolor{black}{#1}}
\newcommand{\xl}[1]{\textcolor{black}{#1}}

\usepackage[normalem]{ulem}
\usepackage{upgreek}

\begin{document}
\title{Band flipping and bandgap closing in a photonic crystal ring and its applications}

\author{Xiyuan Lu}\email{xnl9@umd.edu}
\affiliation{Microsystems and Nanotechnology Division, Physical Measurement Laboratory, National Institute of Standards and Technology, Gaithersburg, MD 20899, USA}
\affiliation{Joint Quantum Institute, NIST/University of Maryland, College Park, MD 20742, USA}

\author{Ashish Chanana}
\affiliation{Microsystems and Nanotechnology Division, Physical Measurement Laboratory, National Institute of Standards and Technology, Gaithersburg, MD 20899, USA}

\author{Yi Sun}
\affiliation{Microsystems and Nanotechnology Division, Physical Measurement Laboratory, National Institute of Standards and Technology, Gaithersburg, MD 20899, USA}
\affiliation{Joint Quantum Institute, NIST/University of Maryland, College Park, MD 20742, USA}

\author{Andrew McClung}
\affiliation{Department of Electrical and Computer Engineering, University of Massachusetts Amherst, Amherst, MA 01003, USA}

\author{Marcelo Davanco}
\affiliation{Microsystems and Nanotechnology Division, Physical Measurement Laboratory, National Institute of Standards and Technology, Gaithersburg, MD 20899, USA}

\author{Kartik Srinivasan}\email{kartik.srinivasan@nist.gov}
\affiliation{Microsystems and Nanotechnology Division, Physical Measurement Laboratory, National Institute of Standards and Technology, Gaithersburg, MD 20899, USA}
\affiliation{Joint Quantum Institute, NIST/University of Maryland, College Park, MD 20742, USA}

\date{\today}

\begin{abstract}
      \noindent The size of the bandgap in a photonic crystal ring is typically intuitively considered to monotonically grow as the modulation amplitude of the grating increases, causing increasingly large frequency splittings between the 'dielectric' and 'air' bands. In contrast, here we report that as the modulation amplitude in a photonic crystal ring increases, the bandgap does not simply increase monotonically. Instead, after the initial increase, the bandgap closes and then reopens again with the dielectric band and the air bands flipped in energy. The air and dielectric band edges are degenerate at the bandgap closing point. We demonstrate this behavior experimentally in silicon nitride photonic crystal microrings, where we show that the bandgap is closed to within the linewidth of the optical cavity mode, whose quality factor remains unperturbed with a value $\approx$ 1$\times$10$^6$ (linewidth $<$1.5~pm). Moreover, through finite-element simulations, we show that such bandgap closing and band flipping phenomena exist in a variety of photonic crystal rings with varying units cell geometries and cladding layers. At the bandgap closing point, the two standing wave modes with a degenerate frequency are particularly promising for single-frequency lasing applications. Along this line, we propose a compact self-injection locking scheme that integrates many core functionalities in one photonic crystal ring. Additionally, the single-frequency lasing might be applicable to DFB lasers to increase their manufacturing yield.
\end{abstract}

\maketitle

\section{Introduction}
\noindent A grating is one kind of photonic crystal structure, with periodic modulation of the dielectric function in one dimension~\cite{Joannopoulos_book, Johnson_OE_2001_Block}. The modulation can have various shapes, including but not limited to square, triangular, and sinusoidal profiles. In such a periodic grating, there often exist `dielectric' band and `air' band modes, which are two standing-wave modes whose dominant electric field amplitude maxima (i.e., field antinodes) overlap with the dielectric or the air part of the structure. Because the dielectric has a higher refractive index than air, the dielectric mode typically can accommodate a larger wavelength, so that it is at a lower frequency than the air mode. However, a counter-intuitive bandgap closure effect, even for large modulation values, has been observed in photonic wire Bragg gratings, and is predicted to generally exist in various periodic grating structures~\cite{Gnan_OE_2009_Closure, Lee_OL_2019_Essential}.

This one-dimensional grating structure can be implemented into a microcavity~\cite{Cai_Science_2012_Integrated,Lee_OL_2012_Slow, Feng_Science_2014_Single,Lu_APL_2014_Selective, Arbabi_OE_2015_Grating,Yu_NatPhoton_2021_Spontaneous, Lu_PhotonRes_2020_Universal, Lu_NatPhoton_2022_High, lu_highly-twisted_2023}, such as a microdisk or a microring, with various configurations in terms of modulation period and amplitude) resulting in orbital angular momentum (OAM) emission~\cite{Cai_Science_2012_Integrated, lu_highly-twisted_2023}, slow light and defect mode localization~\cite{Lee_OL_2012_Slow, Lu_NatPhoton_2022_High}, single mode lasing~\cite{Feng_Science_2014_Single, Arbabi_OE_2015_Grating}, and spontaneous pulse formation~\cite{Yu_NatPhoton_2021_Spontaneous}. In the grating structure with small amplitudes (a few nanometers to several tens of nanometers), the physics of OAM emission, where light is ejected vertically by the grating, has recently been linked with that of `selective mode splitting' (SMS)~\cite{Lu_APL_2014_Selective}, where light is reflected backwards by the grating, enabling quantitative understanding of OAM emission in a microcavity~\cite{lu_highly-twisted_2023}. 

\begin{figure*}[t!]
\centering\includegraphics[width=0.9\linewidth]{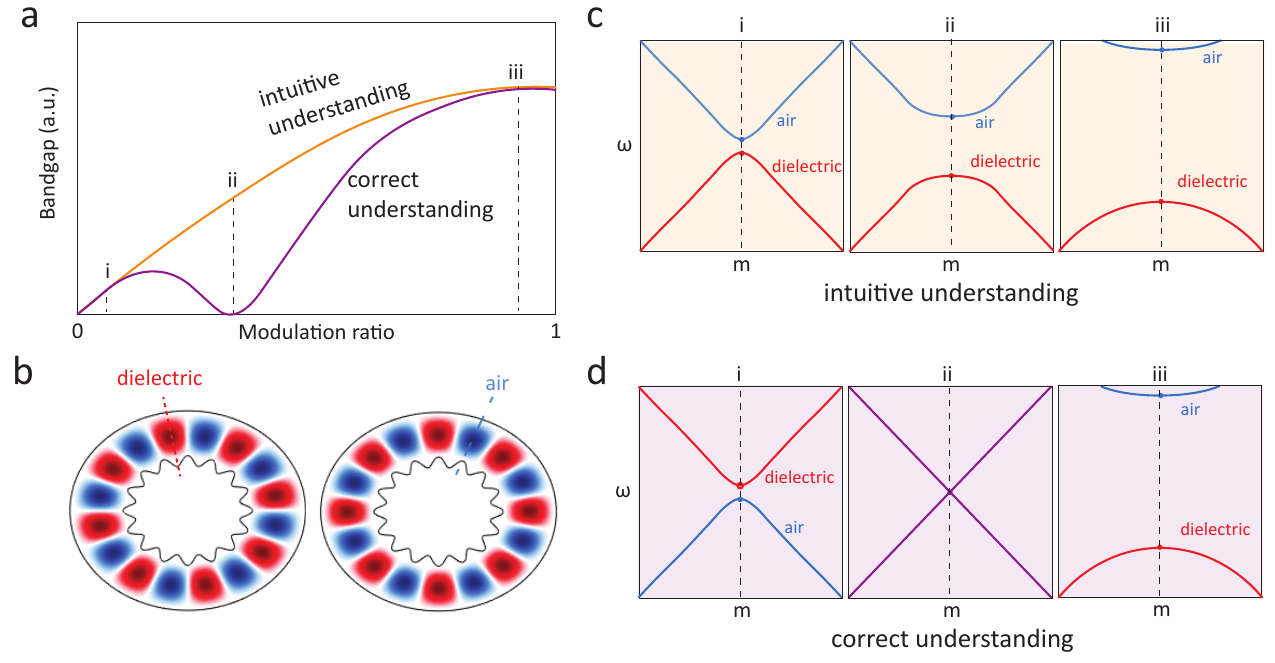}
\caption{\textbf{Band flipping and bandgap closing in a photonic crystal ring.} \textbf{a,} Intuitive understanding, typified by a perturbation theory analysis, suggests that the bandgap of a photonic crystal structure monotonically depends on the modulation ratio (for a PhCR, the ratio of the corrugation size to the ring width). In this work, we report the flipping of the dielectric and air bands, and also the complete closure of the bandgap. \textbf{b,} The bandgap is defined by the difference of the frequencies of the air and dielectric band-edge modes, which are typically taken to be two standing-wave modes with their maximum dominant electric field amplitude located at the widest and narrowest part of the ring, respectively. \textbf{c,} The two limiting cases of small (i) and large (iii) modulation correspond to experimentally studied regimes of selective mode splitting~\cite{Lu_APL_2014_Selective} and the `microgear' photonic crystal ring~\cite{Lu_NatPhoton_2022_High}, respectively. Within those regimes, it is typically assumed that the air band always has higher frequency than the dielectric band, and that the bandgap thus monotonically increases with the modulation ratio, as illustrated by (i-iii). \textbf{d,} In this work, we clarify the picture of band flipping and bandgap closing in a photonic crystal ring. For small modulation (i), the dielectric band actually has higher frequency than the air band. As a result, there is a guaranteed band flipping (i-iii). At the flipping point (ii), the bandgap closes, with the dielectric and air band-edge mode degenerate in frequency.}
\label{Fig1}
\end{figure*}

A perturbation theory for Maxwell’s equations has been developed to calculate the change of resonance frequency of a given electromagnetic mode with shifting material boundaries~\cite{Johnson_PRE_2002_Perturbation}. Following this theory, it is straightforward to calculate the mode frequencies of the dielectric or air mode induced by the grating in a microdisk~\cite{Lu_APL_2014_Selective} or a microring~\cite{Lu_PhotonRes_2020_Universal}. The calculated air mode always has higher frequency than the dielectric mode, which seems quite natural based on typical results observed in such photonic crystal ring (PhCR) devices~\cite{Lee_OL_2012_Slow, Lu_NatPhoton_2022_High}. This intuitive understanding of the PhCR is illustrated by the yellow curve in Fig.~\ref{Fig1}(a). As shown in Fig.~\ref{Fig1}(b), the air mode experiences a narrower ring width (and lower effective index) more, so it seems natural that it should have a shorter resonance wavelength and thus higher frequency, compared to the dielectric mode that experiences a wider ring width. Based on such intuition, the overall physical picture of a PhCR, from the perturbative~\cite{Lu_APL_2014_Selective} [Fig.~\ref{Fig1}(c)(i)] to the non-perturbative regime~\cite{Lu_NatPhoton_2022_High} [Fig.~\ref{Fig1}(c)(iii)], would seem to be understandable as a simple combination and interpolation between these two endpoints, as illustrated in Fig.~\ref{Fig1}(c).

There have been a few reports indicating that bandgap closing behavior, similar to that previously described for the linear one-dimensional gratings~\cite{Gnan_OE_2009_Closure, Lee_OL_2019_Essential}, is also impacting PhCR studies. In particular, it was observed in simulations in Ref.~\cite{Lu_NatPhoton_2022_High}, reported as an adverse effect in Ref.~\cite{Lucas_NatPhoton_2023}, and cited as a major complicating factor in Fourier synthesis dispersion engineering based on multi-frequency gratings in Ref.~\cite{moille_fourier_2023}. However, no direct experimental observation of the bandgap closing has thus far been seen in PhCR studies. In this paper, we report detailed experimental and simulation data validating band flipping and bandgap closing in PhCRs. In particular, we argue that the perturbation theory~\cite{Johnson_PRE_2002_Perturbation} is not applicable to PhCRs with smaller modulations~\cite{Lu_APL_2014_Selective}, and the dielectric band and air bands flip as modulation amplitude is increased, as illustrated in Fig.~\ref{Fig1}(d)(i). As a result, the band physics also needs to be revised to the purple curve in Fig.~\ref{Fig1}(a), so that band flipping [Fig.~\ref{Fig1}(d)(i-iii)] and bandgap closing [Fig.~\ref{Fig1}(d)(ii)] is guaranteed.

To support our argument, we provide experimental measurement verifying such band flipping and bandgap closing phenomena. The experimental results are in close agreement with full three-dimensional finite-element-method simulation of a unit photonic crystal cell. We further carry out simulations for various device geometries, and conclude that this band flipping and bandgap closing behavior exists in many geometries, regardless of symmetric or asymmetric cladding, inside or outside modulation, as well as different modulation shapes.

In particular, around the bandgap closing point, the device has dielectric and air modes at almost degenerate frequencies, both with high optical quality factors close to a million. These modes retain their standing wave nature, as expected from the strong backscattering provided by the grating, and confirmed through measurements of the reflection spectrum from the cavity. As a result, a PhCR device operating near the bandgap closing point can provide narrow bandwidth backscattering for an incoming laser signal, which is particularly useful for linewidth narrowing in lasing applications. The high optical quality factors also make the device suitable for efficient nonlinear optical processes. Along these lines, we propose an ultra-compact injection locking scheme, which can integrate a high-quality-factor cavity, mode selection and filtering, and single-frequency signal reflection, all in the footprint of a single photonic crystal microring.

\medskip
\section{Experimental proof of band flipping and bandgap closing}
\begin{figure*}[t!]
\centering\includegraphics[width=0.95\linewidth]{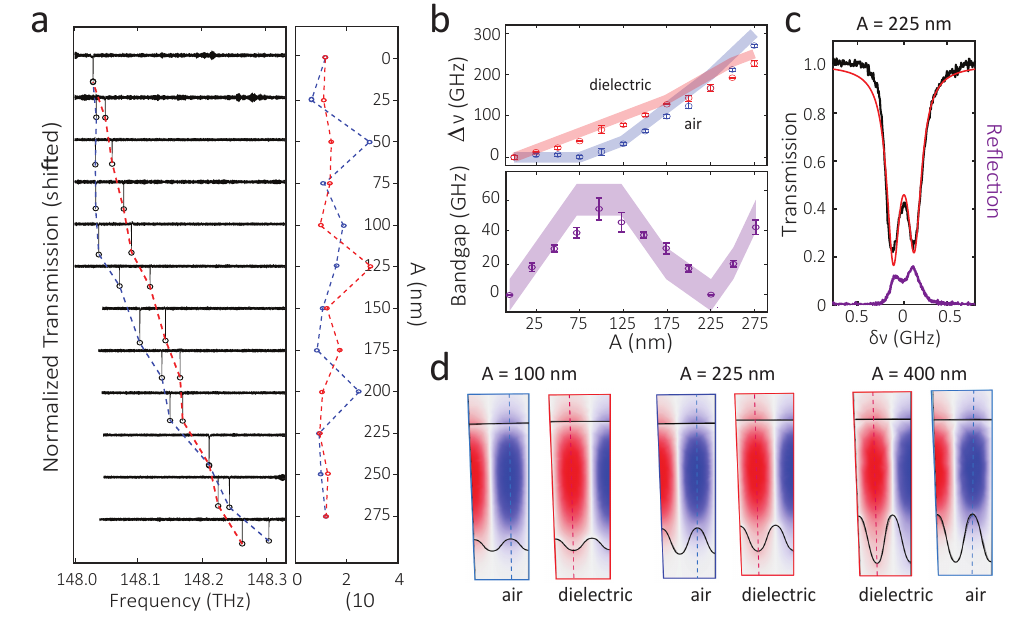}
\caption{\textbf{Experimental verification of the band flipping and bandgap closing.} \textbf{a,} Series of cavity transmission spectra illustrating the crossing of the dielectric (red dashed line) and air (blue dashed line) band-edge frequencies of the photonic crystal ring. 
The loaded optical quality factor ($Q_t$) for each modulation amplitude (right) shows no observable degradation due to the modulation, and remains within (1.4$\pm$0.6)$\times10^6$, where the uncertainty indicates the one-standard-deviation range of $Q_t$ as fit for all twelve devices. \textbf{b,} We plot the frequency offsets with respect to the control device without modulation for the dielectric and air band-edge modes in the top graph, and their frequency difference (i.e., the bandgap) in the bottom graph, clearly showing a bandgap closing and reopening trend. 
\textbf{c,} We show the band-closing device (A$= 225$~nm) with a transmission spectrum (black) and reflection spectrum (purple) containing both dielectric and air band-edge modes with 3~pm separation between the resonance centers, within the full-width at half-maximum of resonance linewidth. The fit (shown in red) for the transmission curve shows the intrinsic $Q$ to be (9.2$\pm$0.1)$\times10^5$, where the uncertainty indicates the 95\% confidence range of the fitting. 
\textbf{d,} The dominant electric field (along the radial direction) of the dielectric and air band-edge modes in devices with three modulation amplitudes, simulated by the finite-element method.
}
\label{Fig2}
\end{figure*}

\noindent We demonstrate this band flipping and bandgap closing phenomenon in high-Q photonic crystal microring (PhCR) devices in the silicon nitride nanophotonics platform~\cite{Moss_NatPhoton_2013_New}. The silicon nitride platform has been successful in application to frequency combs~\cite{Gaeta_NatPhoton_2019_Photonic}, optical parametric oscillation~\cite{Lu_Optica_2019_Milliwatt}, and recently injection locking~\cite{Kondratiev_FrontPhys_2023}. PhCR devices with small modulation amplitudes ($<$ 100~nm) in this platform have been used for frequency combs~\cite{Yu_NatPhoton_2021_Spontaneous, Lucas_NatPhoton_2023, moille_fourier_2023} and optical parametric oscillation~\cite{Lu_OL_2022_Kerr, Jennifer_Optica_2022_Tunable} to provide additional dispersion engineering functionalities. PhCR devices with large amplitudes (several hundred nanometers) have been studied for slow light and defect localization~\cite{Lu_NatPhoton_2022_High}.

Here we report the band flipping and bandgap closing phenomenon in a set of twelve PhCR devices with the inside ring radius modulated with a sinusoidal shape of differing amplitude ($A$). The microring devices are fabricated in 600~nm Si$_3$N$_4$ on 3~$\mu$m thick thermal SiO$_2$ grown on a Si wafer, using an electron-beam lithography nanofabrication method~\cite{Lu_NatPhys_2019}. The PhCR devices have an outer ring radius of 25~$\mu$m and an average ring width of 2~$\mu$m. The number of modulation periods within the ring circumference is 228 throughout all rings, and is designed to create a mode splitting for the mode with an azimuthal mode number of $m$ = 124. 

The fundamental transverse-electric-like mode with this targeted mode number has a resonance frequency around 148~THz for these PhCR devices, as shown in Fig.~\ref{Fig2}(a), with $A$ from 0~nm (top) to 275~nm (bottom). The mode splitting, namely, the frequency separation between the two split modes, increases for $A$ from 0~nm to 100~nm, and then decreases and nears zero at $A$ of 225~nm, and then reopens again. The total loaded quality factor ($Q_t$) remains high, around $1\times10^6$, throughout these devices, as shown in the right panel of Fig.~\ref{Fig2}(a). The mode splittings are typically so large with respect to the mode linewidths that each mode can be fit to a simple Lorentzian model, with $1/Q_t = 1/Q_0 + 1/Q_c$, where $Q_0$ and $Q_c$ are intrinsic and coupling quality factors. At $A$=225~nm (the bandgap closing point), the mode is not a fully resolved doublet (Fig.~\ref{Fig2}(c), and a standard backscattering-induced modal coupling model~\cite{Borselli_OE_2005} is used to extract $Q_t$.


The mode splitting reflects the bandgap of the PhCR devices, but it is difficult to identify which one is the dielectric and which is the air mode, as illustrated in Fig.~\ref{Fig1}(b). In principle, it may be possible to tell from high resolution imaging of the optical field, though this is especially challenging considering the 2~$\mu$m wavelength of the modes. Our mode identification is based on comparison of simulation results with the measured shifts of frequencies to those in the control ($A$ = 0~nm) devices in Fig.~\ref{Fig2}(b) (we also compare the bandgap evolution to simulation). The error bars are one-standard-deviation uncertainty of the laser frequency from multiple scans. The shaded areas are based on three-dimensional finite-element-method simulations, with their width representing the one-standard deviation uncertainty based on resolution of the simulation. We see that the experimental result is in good agreement with simulation, and that the dielectric mode and air mode can be unambiguously identified. It is interesting to note that the initial flatness of the air mode (small shift with $A$ relative to the unperturbed cavity mode position) in the measured data is captured by the simulation too, giving us high confidence to trust the simulation model and results. We conclude from Fig.~\ref{Fig2}(b) that the dielectric and air band has a flip in the PhCR device set we tested, following the illustration of band physics we proposed (purple) in Fig.~\ref{Fig1}. Figure~\ref{Fig2}(c) shows the zoom-in transmission for the device with $A$ = 225~nm, where the bandgap is closed to a level on par with the cavity linewidth. The intrinsic optical quality factor is fit (shown in red) to be $(9.24\pm0.05)\times10^5$, with the uncertainty indicating the $95~\%$ confidence interval of the nonlinear doublet fitting. It is difficult to tell which of the split modes is the dielectric mode and which is the air mode. The curve in purple in Fig.~\ref{Fig2}(c) shows the corresponding reflection signal measured at the input waveguide, where the input power was calibrated to give the absolute reflection percentage. The curve shows a reflection of $20.5 ~\%$ of the input power at the resonance where bandgap closing was observed. The presence of this considerable level of reflection signal suggests that the modes are still standing waves in nature, even though they exhibit a very small frequency splitting compared to their linewidths. This phenomenon of `reflection without splitting' has great significance to single-frequency lasing, which we will discuss in Section 4. To the best of our understanding, this is the first experimental validation of the complete closure of the bandgap in PhCRs, which was previously suggested in simulation~\cite{Lu_NatPhoton_2022_High,moille_fourier_2023} and a measurement of the bandgap starting to shrink with increased modulation amplitude was presented in a recent work~\cite{Lucas_NatPhoton_2023}.

We plot in Fig.~\ref{Fig2}(d) six simulated mode profiles from a top view, that is, the dominant electric field in intensity and phase of the dielectric and air band-edge modes before, at, and after the band closing/flipping. Here red and blue indicate the positive or negative direction of the electric field, and the darkness of each color indicates the field amplitude. For the larger modulation $A$ = 400~nm, the dielectric mode has a larger mode volume than the air mode, which aligns with the common understanding in gratings or photonic crystals, and is consistent with the mode being at longer wavelength/lower frequency. At $A$ = 225~nm, the dielectric mode and the air mode have similar mode volume (consistent with the frequencies being similar), though their antinodes are sitting in the widest and narrowest part of the ring, respectively. It seems that considering the behavior of the field across the entirety of the unit cell is important to understand the frequency of the two modes, and that merely considering the location of the antinodes with respect to the modulation can be misleading.




\begin{figure*}[t!]
\centering\includegraphics[width=0.9\linewidth]{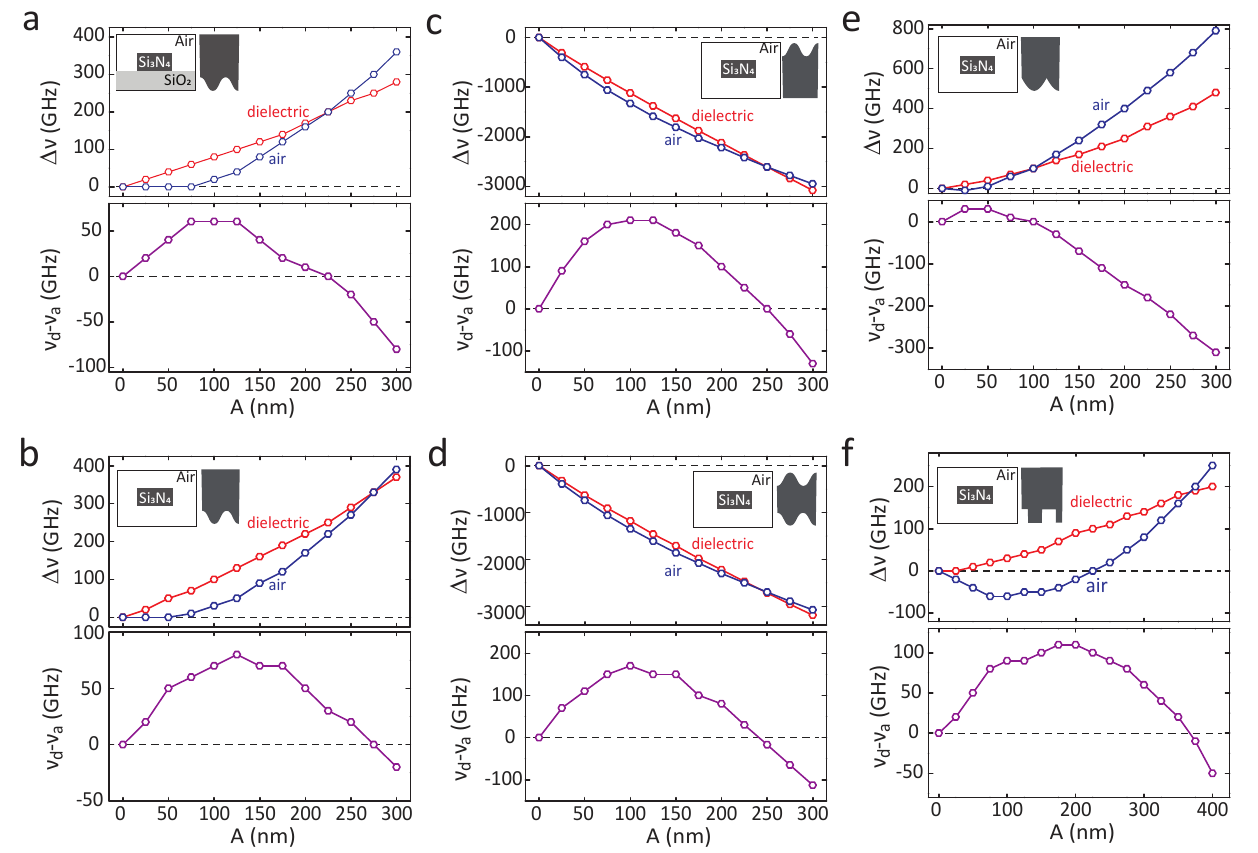}
\caption{\textbf{Simulation results showing the general existence of the band flipping and bandgap closing behavior.} \textbf{a,} Simulation result for the experimentally studied system from Fig.~\ref{Fig2}. The simulation data is shown without error bars (amplitude is $\approx$~10~GHz, as given by the simulation resolution). The left inset shows the cross-section view of the silicon nitride (Si$_3$N$_4$) core with top air cladding and bottom silicon dioxide (SiO$_2$) substrate. The right inset shows the top view of 1.5 unit cells of the photonic crystal ring, with the inside ring radius modulated in a sinusoidal profile. We note that the flatness of the air band-edge mode frequencies and the closing of the bandgap at 225~nm closely resembles the measurements shown earlier. \textbf{b,} Similar band flipping and bandgap closing behavior is observed for a geometry with symmetric top and bottom air cladding, and an inside sinusoidal modulation. \textbf{c-d,} Similar behavior is obtained for an outside modulation only (c), or when both inside and outside modulation (d) is used. In these two cases, the resonance frequencies decrease with increased amplitudes. \textbf{e-f,} Similar behavior is obtained for a folded-sinusoidal modulation profile (e) or a square modulation profile (f) for the inside ring radius.}
\label{Fig3}
\end{figure*}

\section{Simulation verification of the general existence of bandgap closing and band flipping}
\noindent In the previous section, we show that the three-dimensional simulation accurately predicts the experimental results and it also captures the behavior of the flatness of the air mode for small modulation, which gives us good confidence in the simulation model. In this section, we extend this simulation model to various device geometries to show that the band flipping and bandgap closing phenomenon generally exists in PhCRs.

Our previous simulation result is re-plotted in Fig.~\ref{Fig3}(a), where the PhCR has a silicon dioxide (SiO$_2$) substrate. In Fig.~\ref{Fig3}(b), we present the results for a fully air-clad system (i.e., a symmetric cladding). The bandgap closing point shifts from $A=$~225~nm to $A=$~275~nm, yet the overall trend stays the same. In Fig.~\ref{Fig3}(c) the PhCR has an outside ring radius modulation (Fig.~\ref{Fig3}(a)-(b) had an inner ring radius modulation), while in Fig.~\ref{Fig3}(d) both the inside and outside ring radius are modulated. In both cases, the bandgap closing point is around $A=$250~nm. The frequencies of both `dielectric' and `air' modes shift to smaller frequencies in a similar way, as the outside ring radius modulation has a greater effect on the whispering gallery modes than the inside ring radius modulation (a natural consequence of the radial location of the whispering gallery mode within the ring cross-section). We also consider `folded'-sinusoidal shape PhCRs~\cite{Lu_NatPhoton_2022_High} in Fig.~\ref{Fig3}(e) and square-grating PhCRs in Fig.~\ref{Fig3}(f). Both geometries have different bandgap closing points, but the overall phenomenon is similar. For these six simulated geometries, we find that the dielectric mode always shifts its frequency in a nearly linear fashion versus the modulation amplitude $A$, while the `air' mode has a quadratic dependence on the modulation amplitude. Further investigation is needed to understand this behavior quantitatively.

\section{Applications}
\noindent Our findings clarify the band physics of PhCRs, which is important in applications that need accurate \ks{modal} frequency control, for example, \ks{in realizing frequency matching for parametric nonlinear optics including frequency comb generation} and optical parametric oscillation. Of particular usefulness is \ks{the modal degeneracy} at the bandgap closing point for lasing applications. These two degenerate modes offer a reflection signal from the microring to the waveguide, \xl{as shown in Fig.~\ref{Fig2}(c),} \ks{and experimentally we have observed a mode splitting similar to the cavity linewidth. In practice, full degeneracy should be achievable through fine-tuning of $A$, at which point a single reflection peak would be observed, providing a valuable feedback mechanism while avoiding dual-frequency competition.} Along this line, we propose a compact scheme for single-frequency injection locking, and further use of this bandgap closing for single-frequency distributed feedback lasers. 

\begin{figure*}[t!]
\centering\includegraphics[width=0.90\linewidth]{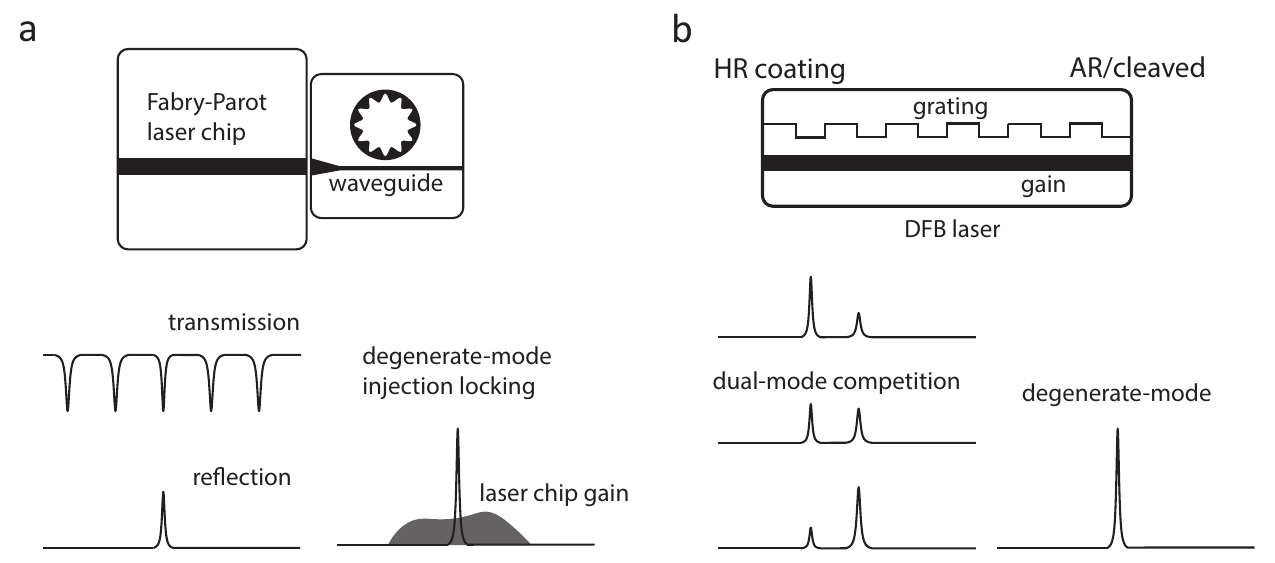}
\caption{\textbf{Application schemes for injection locking of Fabry-Perot lasers and mode selection in distributed feedback (DFB) lasers.} \textbf{a,} In injection locking using a microring resonator, \ks{Rayleigh backscattering due to random ring surface roughness, an introduced grating to create a PhCR structure, or a loop mirror that returns the ring-filtered spectrum in the backward direction} is typically needed to provide feedback to the Fabry-Perot laser. An internal grating in the ring has potential advantages in design, robustness, and compactness, particularly considering the selection of one cavity mode, but can suffer dual-mode competition in conventional mode splitting devices, due to feedback from both split modes. Having degenerate standing-wave modes can potentially eliminate such mode competition and achieve injection locking in an ultra-compact footprint. \textbf{b,} DFB lasers often have a cleaved facet for the laser output for convenience in fabrication, but their yield is limited by the random phase of the cleaved facet that leads to competition between two modes supported by the distributed grating, which typically has a square profile. This problem may be bypassed when adjusting the parameters so that two modes are degenerate in frequency, similar to the simulation result shown in Fig.~\ref{Fig3}(f).}
\label{Fig4}
\end{figure*}

Injection locking is a method to turn a multi-mode laser with a broad and noisy frequency spectrum, for example, a Fabry-Perot laser, to a single-mode laser with a single, low-noise frequency output~\cite{Kondratiev_FrontPhys_2023}. Achieving injection locking requires a spectrally narrow reflection signal with controllable phase. Typically, this simultaneously requires a high-quality-factor cavity, a mechanism for single mode selection or filtering, and a reflection signal with controlled phase. In our scheme, these three functions can be elegantly solved with one photonic crystal microring. In particular, at the bandgap closing point, the device produces a narrow-band reflection signal at only one single frequency, in comparison to reflection signals at two frequencies at other operation points. Phase control can be realized through, for example, incorporation of a thermal phase-shifter on the input waveguide. Our proposed approach provides some simplification in number of components in comparison to those using microring resonators and a loop mirror for feedback. Alternately, a PhCR operating away from the bandgap closing point can be used.  Indeed, this has recently been demonstrated in the context of frequency comb generation~\cite{Ulanov_arXiv_2023}.  We anticipate that the single degenerate mode scheme proposed here might provide additional benefits in such applications.

Another application case is distributed-feedback lasing, where dual-mode competition has been identified as an issue in laser manufacturing that reduces the yield~\cite{Buus_ElectronLett_1985_Mode, Mccall_JQE_1985_An}. This dual-mode competition is attributed to the random phase created by the cleaved facet~\cite{Buus_ElectronLett_1985_Mode}. While putting a phase slip with the correct size and placement can provide mode discrimination~\cite{Mccall_JQE_1985_An} in the distributed grating, the use of uniform gratings and removal of devices exhibiting dual-mode competition persists. Operation at the bandgap closing point presents one potential alternative to a phase-slip grating. Ultimately, the utility of this approach depends on the extent to which the grating fabrication is practical in design, yield, and cost. To that end, it seems likely that the gratings prescribed in this work can be realized through photolithography rather than the electron-beam lithography used here, though careful calibration of the lithographic pattern may be needed to ensure operation at the band closing point.


\medskip
\medskip
\section{Discussion}
\noindent We report band flipping and bandgap closing in a photonic crystal ring. This finding adds to the understanding of the physics of the photonic crystal ring, and is crucial for applications when the amount of splitting (size of the bandgap) created is of importance. This can include dispersion engineering for nonlinear optics like optical parametric oscillation~\cite{Lu_OL_2022_Kerr} and Kerr frequency comb~\cite{Yu_NatPhoton_2021_Spontaneous}, as well as the usage of orbital angular momentum emission~\cite{Cai_Science_2012_Integrated,lu_highly-twisted_2023}. We also discuss the potential applications for improving injection locking and DFB lasing.

This work focuses on one category of PhCRs that implement a one-dimensional grating into the microring~\cite{Lee_OL_2012_Slow}, with the advantage of high-quality-factor whispering gallery modes and straightforward coupling to access waveguides~\cite{Lu_NatPhoton_2022_High}. There are other type of photonic crystal rings that can work similarly, including but not limited to, recently demonstrated `slit' and `rod' PhCRs~\cite{Lu_Nanophotonics_2023}. Moreover, there is also another category of photonic crystal microresonator, which is built on a closed-circuit waveguide path within a two-dimensional photonic crystal structures, and is also referred to as a photonic crystal `disk/ring' resonator (PCDR/PCRR)~\cite{Smith_APL_2001_Coupled,Kim_APL_2002_Two, Zhang_OL_2014_High}. These devices have the advantage of smaller mode volumes and more degrees of freedom in design. Going forward, it will be interesting to investigate whether there is band flipping in two-dimensional photonic crystal structures at a small filling fraction, similar to that observed here in one-dimensional photonic crystal microrings.



\medskip
\noindent \textbf{Acknowledgements.} X.L. would like to thank Qingyang Yan for her help with editing. The authors acknowledge Govind P. Agrawal, Vladimir A. Aksyuk, Gregory Moille, and Jordan R. Stone for helpful discussions. 

\medskip
\noindent \textbf{Funding.} This work is supported by the NIST-on-a-chip program and the Maryland Innovation Initiative.

\medskip
\noindent \textbf{Disclosures.} University of Maryland has filed an invention disclosure, with X.L., K.S., and A.M. listed as inventors, related to the work presented in this article.\\

\bibliographystyle{naturemag}
\bibliography{wBG}

\end{document}